# Causal pathway from AMOC to Southern Amazon rainforest indicates stabilising interaction between two climate tipping elements

Annika Högner[1,2,*], Giorgia Di Capua[2], Jonathan F. Donges[2,4,5], Reik V. Donner[2,3], Georg Feulner[2], and Nico Wunderling[2,5,6]

[1] Energy, Climate and Environment Program, International Institute for Applied Systems Analysis (IIASA), Laxenburg, Austria
[2] Potsdam Institute for Climate Impact Research (PIK), Member of the Leibniz Association, Potsdam, Germany
[3] Department of Water, Environment, Construction and Safety, Magdeburg-Stendal University of Applied Sciences, Magdeburg, Germany
[4] Stockholm Resilience Centre, Stockholm University, Stockholm, Sweden
[5] High Meadows Environmental Institute, Princeton University, Princeton, NJ, USA
[6] Center for Critical Computational Studies (C³S), Goethe University Frankfurt, Frankfurt am Main, Germany

* Corresponding authors: Annika Högner

**Email:** hoegner@iiasa.ac.at

**Author Contributions:** A.H. and N.W. designed the research; A.H. conducted the analysis and produced the figures; A.H. led the writing of the paper with input from G.D.C., J.F.D., R.V.D., G.F. and N.W.; N.W. supervised the study.

**Competing Interest Statement:** The authors declare no competing interests.

## Abstract

Declines in resilience have been observed in several climate tipping elements over the past decades, including the Atlantic Meridional Overturning Circulation (AMOC) and the Amazon rainforest (AR). Large-scale nonlinear and possibly irreversible changes in system state, such as AMOC weakening or rainforest-savanna transitions in the Amazon basin, would have severe impacts on ecosystems and human societies worldwide. In order to improve future tipping risk assessments, understanding interactions between tipping elements is crucial. The AMOC is known to influence the Intertropical Convergence Zone, potentially altering precipitation patterns over the AR and affecting its stability. However, AMOC-AR interactions are currently not well understood. Here, we identify a previously unknown stabilising interaction pathway from the AMOC onto the Southern AR, applying an established causal discovery and inference approach to tipping element interactions for the first time. Analysing observational and reanalysis data from 1982-2022, we show that AMOC weakening leads to increased precipitation in the Southern AR during the critical dry season, in line with findings from recent Earth system model experiments. Specifically, we report a 4.8% increase of mean dry season precipitation in the Southern AR for every 1 Sv of AMOC weakening. This finding is consistent across multiple data sources and AMOC strength indices. We show that this stabilising interaction has offset 17% of dry season precipitation decrease in the Southern AR since 1982. Our results demonstrate the potential of causal discovery methods for analysing tipping element interactions based on reanalysis and observational data. By improving the understanding of AMOC-AR interactions, we contribute toward better constraining the risk of potential climate tipping cascades under global warming.

**Main Text**

**INTRODUCTION**

Climate tipping elements are large-scale bi- or multistable subsystems of the Earth system that can display nonlinear shifts in response to small environmental changes, in particular changes in global mean temperature, with potential repercussions onto the Earth system as a whole (Armstrong McKay *et al* 2022). Examples of systems that may be able to display such nonlinear behaviour include the Greenland Ice Sheet, the Antarctic Ice Sheets, the Atlantic Meridional Overturning Circulation (AMOC), and the Amazon rainforest (AR). High uncertainties remain, in regards to the ability of these systems to actually display tipping dynamics under current climatic conditions, their critical drivers, potential spatially diverse behaviors, the respective critical parameters such as thresholds and time scales, and impacts (Armstrong McKay *et al* 2022). The evolution of fast tipping elements such as the AMOC or the AR, with tipping time scales potentially on the order of decades, are of particular policy-relevant concern (Möller *et al* 2024). The possibility that they could tip within this century with severe global impacts cannot be excluded, and early warning signals of destabilisation have been detected for both (Boulton *et al* 2022, van Westen *et al* 2024). Interactions among climate tipping elements can enhance or alleviate this threat (Wunderling *et al* 2021). For delivering robust tipping risk assessments, it is important to understand these interactions, most of which have been assessed as destabilising, triggering self-amplifying feedbacks (Wunderling *et al* 2024). Some interactions are still accompanied by considerable uncertainties, this includes the strength and sign of the interaction between AMOC and AR, so far assessed as unknown (Wunderling *et al* 2024).

The AR is the Earth's largest rainforest ecosystem and contributes to global temperature regulation via carbon storage and net cooling from evapotranspiration. It is home to over 10% of the world's biodiversity (Flores *et al* 2024). Due to increasing stress from warming temperatures, extreme droughts, and deforestation, AR ecosystem health is in decline in many places and parts of the forest have already turned from carbon sink to source (Gatti *et al* 2021). AR stability was recently summarised to critically depend on global mean temperature, mean annual precipitation, dry season length and intensity, and deforestation (Flores *et al* 2024). AR tipping would imply (partial) forest dieback and vegetation changes towards seasonal forest or savanna, threatening the significant ecosystem services it provides (Armstrong McKay *et al* 2022).

The AMOC is driven by deep-water formation from temperature- and salinity-induced density gradients, convection, evaporation, and wind in the subpolar North Atlantic. It redistributes heat from the equator to higher latitudes. Paleoclimatic evidence indicates past multistability with abrupt changes between a strong and a weak AMOC mainly driven by freshwater influx near Greenland from precipitation and ice sheet melting (Lynch-Stieglitz 2017), and modulated by aerosol concentrations (Schleussner and Feulner 2013, Menary *et al* 2020). AMOC collapse would impact temperature and precipitation patterns globally, reduce Northern Hemisphere warming, shift the ITCZ southward (Bellomo and Mehling 2024), and alter monsoon systems with repercussions on the biosphere across the tropics and beyond (Armstrong McKay *et al* 2022, Feulner *et al* 2013). For this reason, we expect a causal influence from a weakening AMOC onto the AR, however, the sign and strength of the interaction are yet unclear (Wunderling *et al* 2024).

Given the different hydrological cycles in the Southern and Northern AR (Marengo 2006), we expect differences in potential respective causal interaction pathways. We here focus on the Southern AR. Earlier Earth system model (ESM) experiments that induced AMOC weakening or collapse through freshwater hosing in the North Atlantic found significant changes in AR precipitation, however, the different studies reported precipitation changes of contradictory signs for the Southern AR (Parsons *et al* 2014, Jackson *et al* 2015). This disagreement was subsequently attributed to biases in modelling the shift of the Intertropical Convergence Zone (ITCZ) (Good *et al* 2022). A more recent study utilising ESM simulations and a conceptual Stommel two-box model found a competing effect between global warming and AMOC weakening, with AMOC weakening potentially counteracting warming-induced decreases of precipitation in the Southern AR (Ciemer *et al* 2021). This competing effect was further explored in a series of ESM experiments, and evidence for a path-dependency was found, where vegetation is more resilient in scenarios with a weak AMOC, particularly in the South-eastern AR (Nian *et al* 2023), unless AR dieback occurs before AMOC weakening. In the latter case, the weakened AMOC was not found to aid forest regeneration. Another series of North Atlantic freshwater hosing experiments exploring the impacts of AMOC collapse on monsoon systems across several ESMs found a pronounced increase in precipitation and a shortened dry-season length in the Southern AR region (Ben-Yami *et al* 2024a). In summary, ESM-based studies seem to converge towards a stabilising effect from a weakening AMOC onto the Southern AR via increasing precipitation. Paleo-evidence from the AR region indicates a long-term southward shift of the ITCZ (Zhang *et al* 2017) during Heinrich stadials that are associated with a weak AMOC, leading to increased vulnerability of Northern Amazon forests (Akabane *et al* 2024) and suggesting a corresponding increase of Southern AR precipitation. However, evidence from Earth observation data supporting this interaction is still lacking. Here, we utilise a mix of reanalysis and observational time series data to fill this gap.

A recent study provided observation-based evidence of teleconnections between tipping elements of the Earth system for the first time, using correlation-based functional network analysis (Liu *et al* 2023). Here, we use causal discovery, an advanced statistical method that allows us to identify an appropriate structural causal model of the interaction mechanism from data. The idea for this originated in Judea Pearl's theory of causality (Pearl 2009b) and was first introduced to the Earth sciences in 2012 (Ebert-Uphoff and Deng 2012). It has since found wide application in the study of atmospheric teleconnections (Kretschmer *et al* 2017, Di Capua *et al* 2020, Samarasinghe *et al* 2020, Di Capua *et al* 2023, Saggioro *et al* 2024). Pearl showed that causal relationships between variables can be derived purely from observational data, without additional measurements or experimental interventions, under a set of conditions (see Text S1) (Pearl 2009a). This is possible, because a causal graph describing the data contains testable assumptions about the conditional (in)dependence structure among the included variables (Pearl 2009a). An established implementation of causal discovery and inference for time series analysis, that also includes time-lagged versions of the variables, is the PCMCI+ algorithm (Runge 2020). It has been used, for example, for the analysis of atmospheric teleconnections (Di Capua *et al* 2020, 2023), for detecting relationships between modes of long-term internal variability of the climate system (Saggioro *et al* 2020), and for the evaluation of ESMs' ability to represent such relationships (Karmouche *et al* 2023).

We here use PCMCI+ to detect causal interaction pathways between the AMOC and the Southern AR. The AMOC strength/variability is represented by an established sea surface temperature (SST) fingerprint (Caesar *et al* 2018). The Southern AR state is characterised by mean precipitation as well as the Normalised Difference Vegetation Index (NDVI) that describes vegetation greenness. Out of a larger range of potential drivers, mediators, and confounders suggested by the literature, we identify the Caribbean Low Level Jet (CLLJ) as a mediator required in the analysis. By adding a data-driven perspective, our findings deepen the understanding of the interaction between AMOC and Southern AR and narrow the uncertainty around its sign and strength, a crucial step towards improved tipping risk assessments.

**METHODS**

**Data.** The analysis was conducted using monthly resolution data from 1982–2022. The precipitation, wind, and SST-based indices are constructed from ERA5 reanalysis data (Hersbach *et al* 2020), provided by the European Centre for Medium-Range Weather Forecasts (ECMWF). The AR vegetation index is derived from the Global Inventory Modeling and Mapping Studies-3rd Generation

V1.2 (GIMMS-3G+) satellite dataset (Pinzon *et al* 2023) that provides NDVI data in bi-weekly resolution. For sensitivity tests, we also construct the AMOC index using COBE-SST2 (Hirahara *et al* 2014), HadCRUT5 (Morice *et al* 2021), HadSST4 (Kennedy *et al* 2019), and NOAA ERSSTv5 (Huang *et al* 2017) SST data (Figure S1). For an additional precipitation index, we use data from the Global Precipitation Climatology Centre (GPCC) by the Deutscher Wetterdienst (DWD). This is an in-situ reanalysis dataset of global land-surface precipitation from 1981–2020, based on rain gauge data from about 86,000 stations world-wide (Schneider *et al* 2022).

**Index selection.** Beside the main indices representing the two tipping elements (AMOC index, mean precipitation index and NDVI), we need to identify the correct mediators, drivers, and confounders to include in our analysis. For this purpose, we compile a larger selection of indices representing physical processes that the literature suggests as potentially relevant (Table 1) (Högner and Wunderling 2025). We conduct causal discovery on a range of preliminary possible networks from these variables, and identify the CLLJ as a relevant mediator. All other variables from the larger selection do not form stable links that contribute to the interaction directly or alter it in significant ways, suggesting that they do not act as significant drivers, mediators, or confounders on the here analysed timescale and hydrological season. Furthermore, we test the robustness of the finally selected causal effect network against the inclusion of some of the here excluded variables, as well as the substitution of indices through related processes (Text S2, Figure S2).

**Index aggregation.** The main analysis uses four indices: the AMOC index, the CLLJ index, and a mean precipitation index, as well as an NDVI for the Southern AR. The AR indices and CLLJ index are constructed as spatial means aggregated across the respective domains (Figure 1, Table 1). The CLLJ is described as the mean zonal wind speed at 925 hPa in the box bounded by 7.5–12.5° N and 85–75°W (Hidalgo *et al* 2015). We denote easterly wind speed as positive. The AR vegetation index is resampled from bi-weekly into monthly resolution by averaging. The AMOC SST index is adapted from Caesar et al. (Caesar *et al* 2018) by subtracting the global mean SST signal from the mean SST signal in the subpolar gyre region, here using monthly resolution. Additionally, we analyse an alternative SST $AMOC_{dipole}$ index (Figure S3) adapted from Pontes & Menviel (Pontes and Menviel 2024) as the difference between SST averaged over a box in the western South Atlantic (60–30°W; 15–35°S) and the subpolar gyre SST, then subtracting twice the global mean SST, which adds a recently proposed polar amplification correction (Ditlevsen and Ditlevsen 2023). All time series are detrended and deseasonalised by subtracting a first order polynomial least squares fit grouped by the respective month. For a description of the aggregation of the indices used in the preliminary causal discovery and robustness checks, see Table 1.

| Index | Variable | Region | Aggregation | Data source |
|---|---|---|---|---|
| **AMOC fingerprint** | sea surface temperature | subpolar gyre region as defined in Caesar et al. (Caesar *et al* 2018) | mean SST anomaly in subpolar gyre region minus mean global SST anomaly, detrended and deseasonalised | ERA5 (ECMWF) (Hersbach *et al* 2020) HadCRUT5 (Met Office) (Morice *et al* 2021) HadISST4 (Met Office) (Kennedy *et al* 2019) COBE-SST2 (JMA) (Hirahara *et al* 2014) ERSSTv5 (NOAA) (Huang *et al* 2017) |
| **Caribbean Low Level Jet (CLLJ)** | zonal wind speed at 925 hPa | 7.5–12.5°N, 85–75°W | spatial mean, detrended and deseasonalised | ERA5 (ECMWF) |
| **Southern AR precipitation (PREC)** | total precipitation | AR basin South of 5°S | spatial mean, detrended and deseasonalised | ERA5 (ECMWF) GPCC (DWD) (Schneider *et al* 2022) |

| **Southern AR vegetation (NDVI)** | Normalised Difference Vegetation Index | AR basin South of 5°S | spatial mean, detrended and deseasonalised | GIMMS-3G+ (Pinzon et al 2023) |
|---|---|---|---|---|
| AMOC$_{dipole}$ index | sea surface temperature | subpolar gyre region (see above) and a Southern Ocean box 15–35°S, 60–30°W | Mean SST anomaly in subpolar gyre region minus mean SST in Southern Ocean box minus twice the global SST anomalies | ERA5 (ECMWF) HadCRUT5 (Met Office) HadISST4 (Met Office) COBE-SST2 (JMA) ERSSTv5 (NOAA) |
| ENSO1+2 | sea surface temperature | 0–10°S, 90–80°W | spatial average SST anomalies | ERA5 (ECMWF) |
| ENSO3.4 | sea surface temperature | 5°N–5°S, 170–120°W | spatial average SST anomalies | ERA5 (ECMWF) |
| ITCZ | precipitation | 15°N–15°S, 35–15°W (Good et al 2008) | latitudinally weighted zonal mean precipitation | ERA5 (ECMWF) |
| North Atlantic SST (NATL) | sea surface temperature | 5–25°N, 70–15°W | Spatial average (Good et al. 2008), detrended and deseasonalised | ERA5 (ECMWF) |
| South Atlantic SST (SATL) | sea surface temperature | 25–5°S, 40°–20°W | Spatial average (Good et al. 2008), detrended and deseasonalised | ERA5 (ECMWF) |
| Atlantic North South Gradient (ANSG) | sea surface temperature | | NATL - SATL difference | ERA5 (ECMWF) |
| South Atlantic Anticyclone (SAA) longitude | sea level pressure | 10–50°S, 60°W–20°E | longitude of maximum pressure centre | Taken pre-aggregated from Gilliland & Keim 2018 (Gilliland and Keim 2018) |
| South Atlantic Anticyclone (SAA) latitude | sea level pressure | 10-50°S, 60°W–20°E | latitude of maximum pressure centre | Taken pre-aggregated from Gilliland & Keim 2018 |
| North Atlantic Oscillation (NAO) | 500-mb height anomalies | Principal Component centred in the subpolar North Atlantic | Rotated Principal Component Analysis | Taken pre-aggregated from Dool et al. 2000 (Dool et al 2000) |

**Table 1. Indices used in the causal analysis.** All indices utilised in the analysis, the underlying variable and region on which they are constructed, respective method of aggregation, and data sources. The indices used in the main analysis are printed in boldface, all other indices were used in the preliminary analysis and robustness testing.

**Causal discovery and inference.** The analysis is conducted with the *tigramite v5.2* python package (Runge *et al* 2023) using the PCMCI+ algorithm, which iteratively tests conditional independence between time series variables, including time lagged versions, here using a linear partial correlation test (Runge *et al* 2019). The strength of the causal effects (CE) of the links found in the causal discovery is determined using multiple linear regression. For a detailed description of the methodology, see Text S1 in the Supplementary Material. All causal graphs shown in the figures display causal links with time lags in months, shaded by CE strength. Blue (red) links indicate a negative (positive) CE, i.e. a change in the driver variable causes a change of the opposite (same) sign in the target variable (see Text S1 for details).

**Causal stationarity analysis.** The causal stationarity analysis was performed on extended data from 1940-2022, excluding the NDVI that is only available from 1982. To evaluate the evolution of the CE over time, we split the data into windows of 40 years length and conduct our analysis in a sliding window approach in the two following ways: (1) We conduct causal discovery on the full length of the data, then prescribe the discovered links, and employ the CE analysis for the prescribed links within each window; (2) we conduct causal discovery and subsequent CE analysis on each window.

**Causal maps.** We finally resolve the Southern Amazon region spatially and evaluate the CE strength on the grid cell level using two time series indices (AMOC, CLLJ), and two fields (PREC, NDVI). The fields are provided as spatially resolved grid cell level time series. We prescribe the links between the variables from the previous causal discovery on the aggregated indices and evaluate the CE strength of those prescribed links. We repeat the analysis for AMOC indices constructed from four other SST datasets and identify areas of high/low agreement between the respective causal maps, defining areas of low agreement as those grid cells, in which less than two of the four causal maps based on an alternative data source for SST find a CE within one or two standard deviations from the CE found for the respective grid cell using ERA5. The reference standard deviation for the CE for each link is taken from the analysis of the variance of each link under bootstrapping.

## RESULTS

**Causal pathway from AMOC index to Southern AR.** We identify a robust network of causal interactions between the AMOC index (Caesar *et al* 2018) and the Southern AR precipitation (PREC) and vegetation greenness (NDVI) during dry season (May-September) (Nobre *et al* 2009) with a mediated link via the CLLJ and a direct link from the AMOC index to Southern AR precipitation (Figure 1) using the PCMCI+ algorithm for causal discovery. The time lags of the links are in the range of 2-4 months. In this causal effect network, negative SST anomalies in the subpolar gyre region that indicate a weakening AMOC (Caesar *et al* 2018, Rahmstorf *et al* 2015) lead to an intensified CLLJ and higher Southern AR precipitation. The intensified CLLJ increases Southern AR precipitation. Higher precipitation increases the NDVI. In summary, we find that a weakening AMOC increases dry season precipitation and NDVI in the Southern AR.

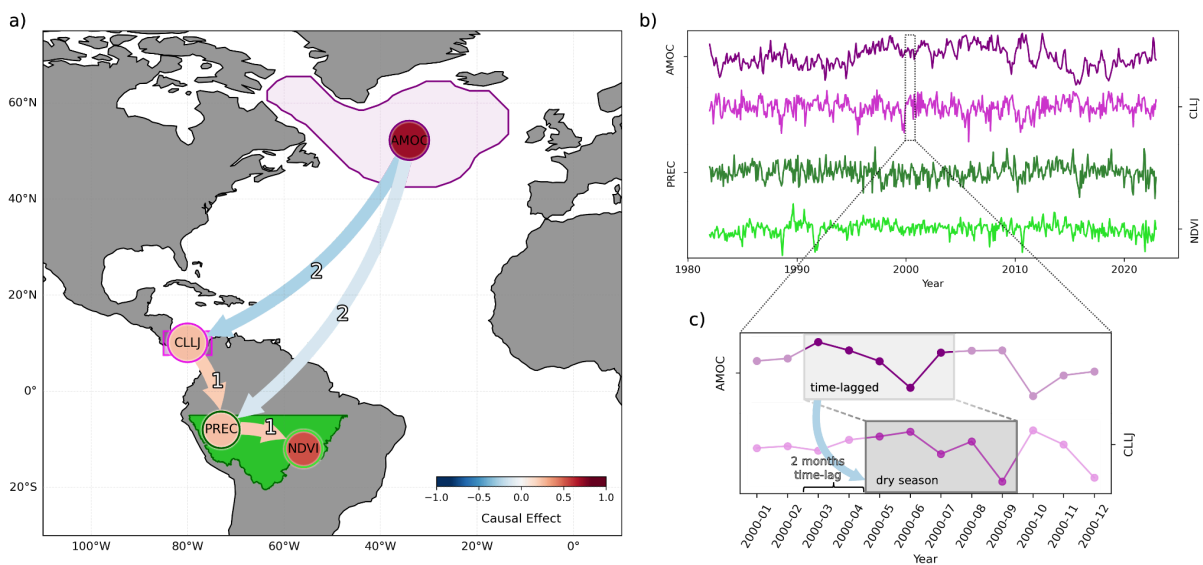

**Figure 1. Causal effect network from Atlantic Meridional Overturning Circulation (AMOC) index to the Southern Amazon rainforest (AR),** a) displayed with the indices placed in their respective location of aggregation: the AMOC index (purple), the Caribbean Low-Level Jet index (CLLJ, magenta), the Southern AR region (green) represented by mean precipitation (PREC, dark green) and mean Normalised Difference Vegetation Index (NDVI, light green). b) The time series of the indices aggregated across the regions shown in panel a), deseasonalised and detrended, as used in the causal discovery. c) An example of the masking and time lag showing data for the year 2000 for the AMOC→CLLJ link, in which the CLLJ as target variable is evaluated for the dry season (May-September), and the driver variable with a time lag of 2 months (March-July).

We perform a series of sensitivity tests on the causal effect network (Text S2), finding it robust against the inclusion of additional variables (Table 1, Figure S3a,b), against the substitution of variables with related processes (Figure S3c), and against the use of different conditional independence tests in the causal discovery (Figure S4). We repeat the analysis constructing the AMOC index from five different SST data sources (ERA5, HadCRUT5, HadISST4, Cobe-SST2, ERSSTv5) and the PREC index from GPCC data (Figure S5), as well as for an alternative $AMOC_{dipole}$ index (Figures S6,S7). We find that the interaction has the correct sign and consistent order of magnitude of the CE across all data sources and AMOC index combinations, confirming the robustness of our results.

**Causal effect strength.** Having identified the interaction structure from the AMOC to Southern AR, we derive the CE strength using multiple linear regression (Text S1). We find that a decrease of the AMOC fingerprint SST anomaly by 1 standard deviation leads to an increase of 0.27 standard deviations in precipitation and 0.1 standard deviations in NDVI in the Southern AR during the dry season (Figure 2a,c). Translating this back into absolute units from the pre-processed time series and utilising the relationship between subpolar gyre SST and AMOC strength from Caesar et al. (Caesar *et al* 2018), we find a monthly precipitation increase of 3.6 mm for 1 Sv of AMOC weakening. This represents a dry season precipitation increase of 4.8%. For a dry season length of five months, this means a total annual dry season precipitation increase of 18 mm per 1 Sv AMOC weakening. Given the current estimate of 0.46 Sv of AMOC weakening per decade (Pontes and Menviel 2024), this translates to a 33.1 mm precipitation increase over the period 1982-2022 expected from the analysed causal effect network alone. However, due to other effects of global warming, observations from ERA5 precipitation data show a drying trend of 4 mm/year dry season precipitation in the Southern AR for 1982-2022 (Figure S8). This corresponds to a cumulative observed dry season precipitation decrease of 160 mm over these four decades. Our results indicate that without the additional precipitation from the interaction from the weakening AMOC, we would have seen a cumulative drying of over 193 mm by 2022 compared with 1982. Thus, the AMOC→AR interaction has offset this drying trend by about 17%.

In order to evaluate the sensitivity of the CE to small changes in the data, we assess its variance under bootstrapping, omitting five random years of data from the initial time series in 500 iterations, and respectively evaluating the CE of the prescribed links (Figure 2b). We find that all links are constrained within an interquartile range of around 0.03 or lower around their mean CE. All links are robust in regards to their sign.

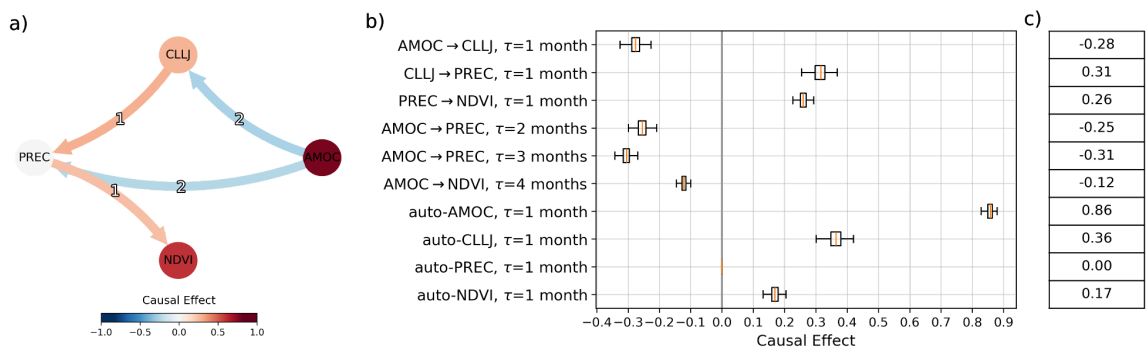

**Figure 2. Causal effect (CE) strength of the AMOC→Southern Amazon rainforest (AR) teleconnection.** a) The network derived with causal discovery using the PCMCI+ algorithm, between AMOC, CLLJ, Southern AR PREC and NDVI. The auto-links constitute self-links of the respective variables. b) CE for the links in the network under 500 iterations of bootstrapping, each randomly omitting 5 years from the data. The whiskers show an added 1.5 IQR to the first and third quartile respectively. c) CE for the links in the network, as listed in b), here evaluated on the full data.

**Causal maps of the Southern AR.** We resolve the Southern Amazon region spatially and evaluate the CE strength for each link in the causal graph between AMOC, CLLJ, PREC, and NDVI that points to one of the AR variables (PREC and NDVI) on the grid cell level (Figure 3). We conduct the analysis for AMOC indices constructed from five different SST datasets (ERA5, HadCRUT5, HadISST4, Cobe-SST2, ERSSTv5). We show the causal maps produced using the ERA5 AMOC index and the distribution of CE for the respective link for all data sources below (Figure 3). All other causal maps, derived with the alternative AMOC index data sources, are available in the Supplementary Material (Figures S9-S12).

The density distributions show agreement on the sign of the respective link across most grid cells, with only tail ends crossing the zero line. While there is some spread in the densities, in particular for the AMOC→PREC links, the overall qualitative agreement across AMOC data sources is good. The central estimates of the distributions of the links are consistent with the CEs previously found in the aggregated analysis (Figure 2) in terms of sign and order of magnitude, although on average a bit weaker when assessed on the grid cell level. We see almost full agreement for the links CLLJ→PREC, AMOC→NDVI, PREC→NDVI across the different SST data sources. We find areas of low agreement based on our threshold definition for the AMOC→PREC links (Figure 3a,b) mainly in the central Southern AR, where the interaction strength is strongest. This is largely due to weaker interaction strength detected in the HadCRUT5 and HadISST4 based AMOC indices (Figures S9,S10). It is worth noting, however, that the agreement on the sign of the interaction across all datasets and links is high.

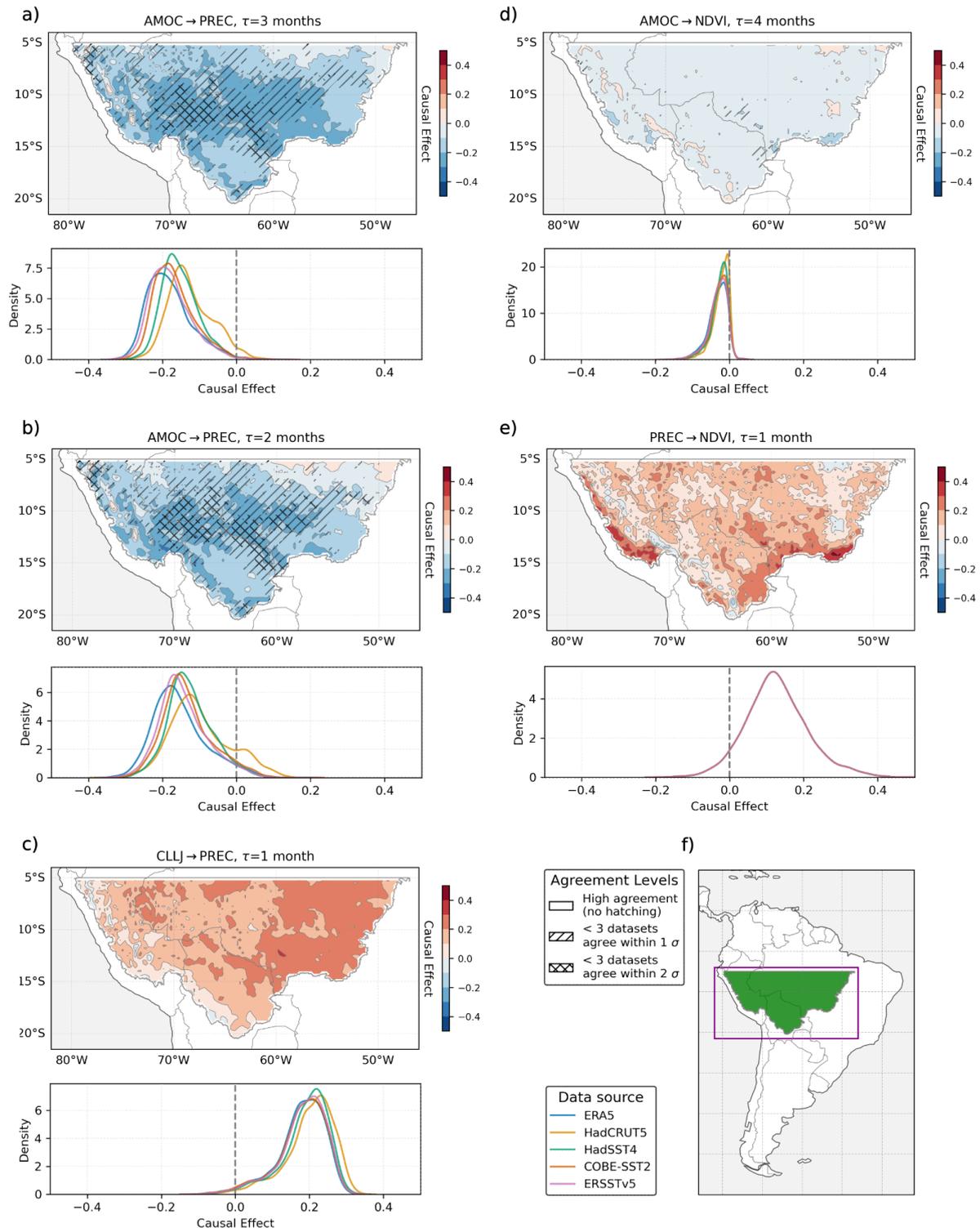

**Figure 3. Causal maps of links into the Southern Amazon rainforest.** The causal effect strength is shown on a grid-cell level for the pathways in the network that point into the Southern Amazon rainforest region, with links to precipitation (PREC) in the left column (a-c) and links to Normalised Difference Vegetation Index (NDVI) in the right column (d, e) for the causal effect network shown in Figure 1a. Hatched regions indicate low agreement between the causal maps (less than three of the five causal maps in agreement). Underneath each causal map, a density plot shows the distribution of the grid cell level CE for the respective map (in blue) and for four maps from alternative data sources for the AMOC index (see Figures S9-S12). f) shows the larger geographical context, with the purple box indicating the bounds of the causal map plots.

**Causal stationarity.** In order to test the validity of the assumption that the causal relationships between the analysed variables are stationary (Text S1), we investigate the evolution of the causal effect network over time for 1940-2022 with a sliding window approach in yearly steps. The network now consists of AMOC, CLLJ, PREC, as NDVI is only available from 1982.

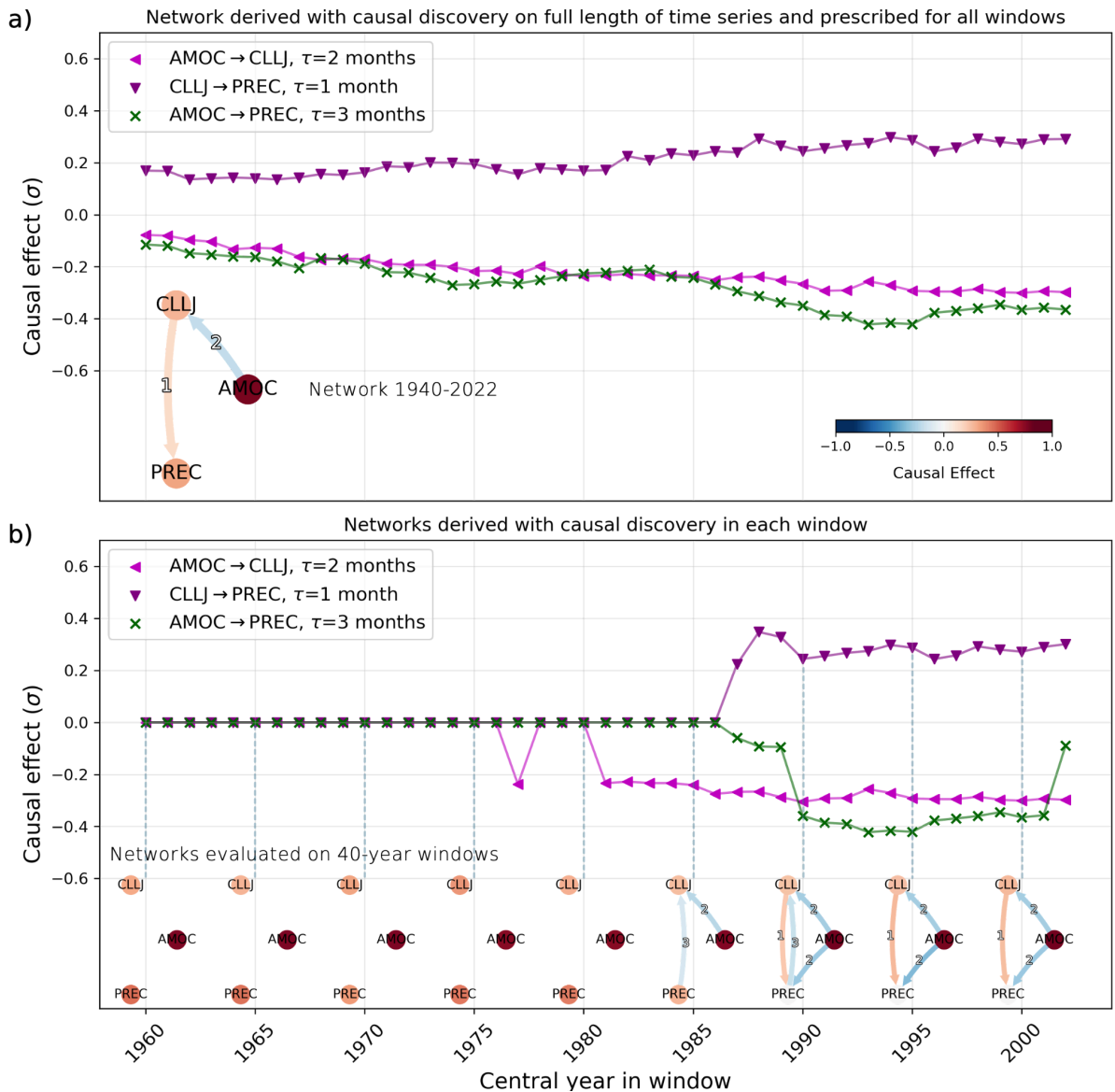

**Figure 4. Causal stationarity analysis.** Evolution of causal effect (CE) for the AMOC, CLLJ, Southern AR PREC network for the years 1940-2022 derived with a sliding window of 40 years length. Each point denotes the CE for the 40-year window centred around the year shown on the *x*-axis. a) The network is derived with causal discovery from the data for the full length of the time series (see inset). The CE over time is derived with multiple linear regression at each time step for links prescribed from this network. b) Causal discovery is conducted for each time step on the data in the respective 40-year window and the CE subsequently derived. We show the evolution of the same links as in a). The causal graphs from the causal discovery for every 5th window (in 5-year steps) are displayed in the inset.

We first conduct causal discovery on the full length of the time series to derive the causal graph (inset Figure 4a), then prescribe its links to derive the CE within each sliding window using multiple linear regression (Figure 4a). We find that all assessed links are consistent in their sign, but gain strength over time. We then perform the full causal discovery on each window separately, with a subsequent multiple linear regression to assess the CE (Figure 4b), evaluating the same links as previously prescribed. This way, we are only able to detect a CE if the links are discovered by PCMCI+. We find that all links are only detected in the later half of the windows. Additional links are found in some windows (inset Figure 4b). Once discoverable, the CE is in a similar range to the CE when links are initially prescribed (Figure 4a). This underlines that although the links are present throughout, they not only become stronger from the 1980s onwards, but also more statistically significant.

**DISCUSSION**

In this study, we identify a previously unknown long-range interaction from an established AMOC SST fingerprint (Caesar *et al* 2018) to the Southern AR analysing observational and reanalysis data with causal discovery and inference methods. We find that AMOC weakening leads to increased dry season precipitation in the Southern AR. This is in agreement with recent ESM based studies that report overall precipitation increases in the Southern AR under a weakening AMOC (Ciemer *et al* 2021, Nian *et al* 2023, Ben-Yami *et al* 2024a). However, we specifically find a precipitation increase in the dry season, which is particularly critical for forest stability. We, thereby, add an observational data-driven perspective to the evolving body of knowledge on this climate tipping element interaction, establishing not only correlation but causality.

Causality here depends on the inclusion of sufficient variables. We have attempted to ensure that sufficiency is met by investigating a range of potential drivers, mediators, and confounders of this interaction. Other variables may still be relevant in AMOC-AR interactions on other time scales, during the wet season and for the Northern AR, which should be investigated in future studies.

Deviations among precipitation data across datasets can be considerable (Sun *et al* 2018, Hassler and Lauer 2021) and reanalysis datasets differ in the observations they are constructed from, in the applied assimilation and bias correction methods, as well as in the utilised interpolation and imputation methods (Calvert 2024). These pre-processing steps frequently prioritise the preservation of mean characteristics of the data over higher-order statistical properties (Ben-Yami *et al* 2024b). We, therefore, repeat our analysis on indices derived from multiple data sources. To strengthen the claim that the AMOC index indeed represents the AMOC, we additionally analyse an alternative $AMOC_{dipole}$ SST index (Pontes and Menviel 2024). We consistently detect a causal interaction pathway from AMOC→Southern AR.

The detected causal effect network starts with the AMOC SST fingerprint in the subpolar North Atlantic, that impacts Southern AR precipitation directly and along a pathway mediated by the CLLJ, subsequently affecting vegetation, with a propagation time of 2-4 months. We hypothesise that the physical interaction pathway unfolds as follows: subpolar North Atlantic SSTs correlate with AMOC strength, cooling when the AMOC weakens (Caesar *et al* 2018, Rahmstorf *et al* 2015), and are linked to the NAO, an atmospheric teleconnection pattern characterised by the exchange of air masses between a low pressure system south-east of Greenland and the North Atlantic Subtropical High (NASH) (Lamb and Peppler 1987). Cooler subpolar North Atlantic SSTs were found to precede positive NAO phases in which the NASH is intensified and shifted to the West (Gastineau and Frankignoul 2015). This strengthens the north-easterly trade winds (Chiang *et al* 2008, Wang and Lee 2007) and leads to a higher moisture influx into Central America (Marengo 2006, Wang and Lee 2007), intensifying the CLLJ (Cerón *et al* 2021, Cook and Vizy 2010). The CLLJ has two annual maxima correlated to the NASH (Cook and Vizy 2010, Wang and Lee 2007), one of which is in July, during the Southern AR dry season. In years of an intensified CLLJ, the ITCZ is shifted southward in June-August (Hidalgo *et al* 2015), leading to higher Southern AR precipitation.

Assessing the evolution of the identified links over time for the reduced network of AMOC, CLLJ, and PREC for the period 1940-2022, we find qualitative continuity of the links and a strengthening of the CEs from the 1980s onwards. Several explanations for this are possible, including 1) changes in data quality given the onset of the satellite era, and 2) decreasing aerosol forcing, which increases drought risk in the AR (Cox *et al* 2008) and contributes to AMOC weakening (Schleussner and Feulner 2013, Hassan *et al* 2021).

A particular pressure on AR stability stems from ongoing deforestation (Flores *et al* 2024). Deforestation has been found to reduce precipitation in the Amazon region (Pires and Costa 2013, Smith *et al* 2023), with particular impact during dry season (Khanna *et al* 2017) and the ability to alter teleconnection patterns (Avissar and Werth 2005). The Southern AR is particularly affected by deforestation (Arias *et al* 2020). Due to data limitations, we did not include deforestation as a confounding variable in our analysis.

Our findings indicate that a weakening AMOC leads to increased precipitation in the Southern AR during the critical dry season, with an effect size of 3.6 mm mean monthly precipitation increase for every 1 Sv of AMOC weakening. This quantitative relationship holds in the current regime, where global warming has reached 1.2 °C (Forster *et al* 2024) relative to pre-industrial global mean temperature and AMOC weakening has been shown to amount to 0.46 Sv per decade since 1950 (Pontes and Menviel 2024). This translates to an additional dry season precipitation of 33.1 mm in 2022 compared to 1982. With an observed dry season precipitation decrease of 160 mm in the same period, our results suggest that without the additional precipitation from the interaction with the AMOC, dry season precipitation would have decreased by 17% more than what is currently observed. The AR is particularly water-limited during the dry season (Gutierrez-Cori *et al* 2021), and dry season intensity was identified as one of the critical drivers of AR stability (Flores *et al* 2024). We, thus, interpret the AMOC→Southern AR interaction as a stabilising interaction between tipping elements.

**CONCLUSION**

We presented evidence for a stabilising interaction from a weakening AMOC onto the Southern AR. Our results contribute to improving climate tipping risk assessments, that so far have largely relied on expert elicitation for estimates to quantify tipping element interactions (Wunderling *et al* 2021, Möller *et al* 2024). We here apply advanced causality-based statistical methods for the first time in the study of tipping element interactions. These methods allow us to use observation and reanalysis data, strengthening the evidence base for the AMOC-Southern AR interaction considerably. This methodology can be applied in the future to investigate other interactions between fast-responding tipping elements, such as the Northern AR, Arctic sea ice, monsoon systems, permafrost (Armstrong McKay *et al* 2022).

Our analysis does not allow a direct extrapolation onto future global warming conditions. Therefore, examining the causal stationarity of the interaction across longer time periods and under more extreme conditions and alternative Earth system states should be subject to future study using ESM simulation data. Follow-up studies with targeted ESM experiments should also assess how deforestation, aerosols, and potential path dependencies affect the interaction identified in this work, and need to investigate multiple time scales of interest.

In conclusion, our findings suggest that without a weakening AMOC, the AR might be losing resilience even more rapidly under ongoing global warming and other anthropogenic pressures such as deforestation. However, despite the stabilising interaction between the two assessed tipping elements, concurrent resilience losses have been observed for both the AMOC (van Westen *et al* 2024) as well as the AR (Boulton *et al* 2022) in recent decades. This implies that other critical drivers of AR stability, such as global warming and deforestation, have destabilising effects that the interaction from the AMOC cannot fully compensate for. Thus, all possible measures to mitigate, and if possible reverse (Schleussner *et al* 2024), additional global warming need to be pursued and further deforestation of the AR needs to be terminated.

**Acknowledgments**
N.W. is grateful for funding from the Pb-Tip project as well as the Center for Critical Computational Studies (C³S). G.D.C. and R.V.D. received financial support from the German Federal Ministry of Education and Research (BMBF) through the JPI Climate/JPI Oceans NextG-Climate Science project ROADMAP (grant no. 01LP2002B). This is ClimTip contribution #32; the ClimTip project has received funding from the European Union's Horizon Europe research and innovation programme under grant agreement No. 101137601: Funded by the European Union. Views and opinions expressed are however those of the author(s) only and do not necessarily reflect those of the European Union or the European Climate, Infrastructure and Environment Executive Agency (CINEA). Neither the European Union nor the granting authority can be held responsible for them. The authors gratefully acknowledge the European Regional Development Fund (ERDF), the German Federal Ministry of Education and Research and the Land Brandenburg for supporting this project by providing resources on the high-performance computer system at the Potsdam Institute for Climate Impact Research.

# SUPPLEMENTARY MATERIAL

**Text S1.**
**Conditions for causal discovery.** Causal relationships between variables without interventional data can be identified under specific assumptions (Runge 2018). In detail, to derive causal information from statistical associations via time series causal graphs, we need to assume that *d*-separation (Pearl 2009a) in the causal graph implies independence of the processes that the variables represent (causal Markov condition) and that independence of the processes implies *d*-separation in the causal graph (faithfulness). We further need to assume that the variables included in the analysis represent all causally relevant processes, i.e. that there are no latent/unobserved confounders (causal sufficiency), and that the causal structure between the variables is time-invariant (causal stationarity). In time series causal discovery, we further assume time-order (cause precedes effect), the data needs sufficient time resolution to avoid undersampling and resolve the processes of interest, and we have to make assumptions about the functional forms of the dependencies between variables when choosing certain conditional association measures for performing the independence test. For a detailed discussion of these conditions and their validity in applications like ours, see (Runge 2018).

**The PCMCI+ algorithm.** The causal discovery and inference analysis is conducted using the *tigramite v5.2* python package (Runge *et al* 2023). It provides an implementation of the PCMCI+ algorithm for causal discovery. This algorithm consists of two iterative steps. First, a Markov discovery step, that uses an adapted version of the PC-algorithm (Spirtes *et al* 2000) (named after its inventors, Peter and Clarke) to identify a preliminary set of causal parents for each variable included in the analysis. Second, a Momentary Conditional Independence (MCI) step that iteratively tests the conditional independencies between the variables using the preliminary sets of parents for conditioning, and that eliminates all spurious links. In the PC step, the preliminary set of possible causal parents for each variable (taken together, the Markov-equivalence class of the causal graph), is obtained by assuming full connectivity between all variables and their time-lagged versions, then iteratively eliminating links between them. This means, we assume that for a variable $V^j_t$ all other variables $V^i_{t-\tau}$ are causal parents, then if a variable is found uncorrelated to $V^j_t$ with a significance level $\alpha_{PC}$, it is removed from the preliminary set of causal parents. For this, all $V^i_{t-\tau} \in S$ are sorted by the strength of their correlation to $V^j_t$, and in every iteration $p$ a subset $S$ is formed of the $p$ strongest links in the preliminary set of parents and conditional independence is tested against this conditioning subset: $V^i_{t-\tau} \perp\!\!\!\perp V^j_t \mid S$. All independent variables are removed from the preliminary set of parents, and the test is repeated for the $p + 1$ strongest links until the algorithm converges (Runge *et al* 2019). In the MCI step, the partial correlation between each possible pair of variables is then estimated through a regression on their combined set of preliminary parents. Different tests are available for the conditional independence testing: Partial Correlation (ParCorr), which is based on bivariate Pearson correlations between variables under the conditioning set, Robust Partial Correlation (RobustParCorr), in which the respective distributions of all variables are Gaussianised prior to the independence testing, and a non-linear Gaussian processes and distance (GPDC) based correlation test. We here use the ParCorr test, however, results for the other two tests are shown in Figure S2. To increase detection power, although at the cost of longer runtime, the PC-step can be omitted, and MCI conducted on all potential lagged links. We use this mode of the algorithm for the robustness test against other SST data sources (Figure S5). For all causality analysis in this study, we use a significance level of $\alpha_{PC}$ = 0.05, a minimum time lag of 0 months, a maximum time lag of 5 months, and adjust the *p*-values with a Benjamini-Hochberg false discovery rate correction (Benjamini and Hochberg 1995).

**Causal effect analysis.** Once the causal graph has been established through causal discovery, the strength of the causal effect (CE) of the links is determined via a multiple linear regression, regressing each variable on its set of causal parents:

$$V^j_t = \sum_i \beta_i V^i_{t-\tau} + \eta^j_t, \qquad (1)$$

where $V^i_{t-\tau}$ are the causal parents of $V^j_t$, $\eta^j_t$ is the residual of the linear regression and the path coefficient $\beta_i$ denotes the CE strength. $\beta_i$ is a dimensionless parameter, and CE is generally given in the units of the respective variable. However, as we standardise all time series before the analysis, we here interpret the CE in terms of standard deviations (Runge *et al* 2023). The assumption of linearity can be made, as the previous causal discovery is conducted using linear conditional independence tests. In order to test the variance of the CE derived in this way, a bootstrapping approach is employed. In the bootstrapping, a fixed number of timesteps is removed randomly from the data, and the CE is evaluated on the remaining bootstrap sample with the links that were previously found in the causal discovery analysis prescribed. This is done iteratively for a number $N$ of bootstrap samples. We use causal maps (Di Capua *et al* 2020) to spatially resolve the CE strength in the Southern AR region, keeping the AMOC and CLLJ indices fixed, while using the grid-cell level time series for the AR variables in the CE analysis.

For the conversion of the CE back to native units of the respective variable, we obtain the standard deviations of the underlying time series prior to standardisation. We do this for the link AMOC→PREC, taking the dry season masking and potential time lags of the respective causal links into account. The conversion from the detrended AMOC fingerprint SST anomalies to actual AMOC strength in units of Sv is done adapting the relationship found in Caesar et al. (Caesar *et al* 2018) in the following way: $\sigma_{AMOC} = 3.8 \frac{Sv}{K} \cdot \sigma_{Index}$, where $\sigma_{AMOC}$ is the standard deviation of the AMOC strength in Sv across the assessed period, and $\sigma_{Index}$ is the standard deviation of the AMOC index in K across the assessed period. The standard deviation for the precipitation is derived from the precipitation time series.

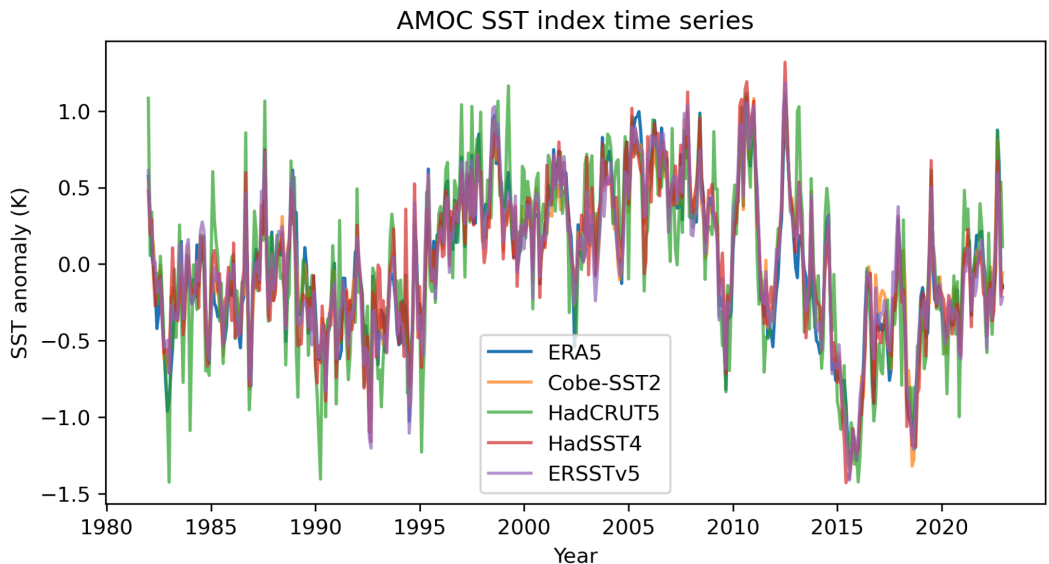

**Figure S1. AMOC SST index from different data sources.** For robustness checks, we use five different reanalysis datasets to construct the AMOC index following Caesar et al. 2018 (Caesar *et al* 2018). We use the ECMWF's ERA5 (Hersbach *et al* 2020), the JMA's COBE-SST2 (Hirahara *et al*

2014), the Met Office's HadCRUT5 (Morice *et al* 2021) and HadSST4 (Kennedy *et al* 2019), and the NOAA's ERSSTv5 (Huang *et al* 2017).

**Text S2.**
We conduct a range of robustness tests for the results of the causal discovery using PCMCI+ (Figures S2, S4-S7). AMOC denotes the AMOC fingerprint constructed using ERA5 SST data. SAA lat and SAA lon the latitude and longitude of the South American Anticyclone, ENSO1+2 the El Niño Southern Oscillation (ENSO) index in the Niño 1+2 region, ENSO3.4 the ENSO index in the Niño 3.4 region, SATL the South Atlantic SST, NATL the North Atlantic SST, ANSG the Atlantic SST North-South gradient, i.e. the difference between NATL and SATL, NAO the North Atlantic Oscillation, and ITCZ an index for the Intertropical Convergence Zone (see Table 1).

We test how the network of AMOC, CLLJ, PREC, NDVI behaves when we include additional variables (Figure S2a,b). No links were found for ENSO1+2, SATL, ANSG, SAA-lat, SAA-lon, and ITCZ (Figure S2a). In all cases, a pathway from the AMOC to the Southern AR variables is recovered, although in some cases the direct AMOC→PREC link is not detected. Including ENSO3.4 recovers a ENSO3.4→CLLJ link (Figure S2b). Including this link slightly weakens the CLLJ→PREC interaction, however, the overall reduction in CE for AMOC→PREC is -0.02, and for AMOC→NDVI there is no change (Table Figure S2b). As the overall effect of including the ENSO3.4 leaves the original network qualitatively intact with only minor effects on CEs and as ENSO3.4 does not link to more than one variable in the network, we do not consider it a relevant confounder. We thereby confirm that the investigated additional variables can be excluded from our analysis.

We test how the network behaves when replacing variables with potentially closely related alternative variables: the AMOC index with the NAO and NATL indices, and the CLLJ with the ITCZ index. Under none of these replacements we recover the original network (Figure S2c), confirming that the variables we have chosen are distinct.

For comparison, we run the causal discovery on the AMOC, CLLJ, PREC, NDVI network using three different conditional independence tests (Figure S4). We find that all three tests recover the CLLJ→NDVI pathway, however, RobustParCorr with a prior Gaussianisation step for all variables does not detect links from the AMOC, and the non-linear Gaussian process regression and distance correlation test (GPDC) does not recover the direct AMOC→PREC link, however, it finds an additional PREC→NDVI link with a time lag of 5 months.

We test multiple combinations of different data sources, using GPCC for precipitation and five different reanalysis datasets for SST (Figure S5), an alternative AMOC$_{dipole}$ index (Figure S6) - shown with ERA5 precipitation, and also analysed with the non-linear GPDC conditional independence test (Figure S7).

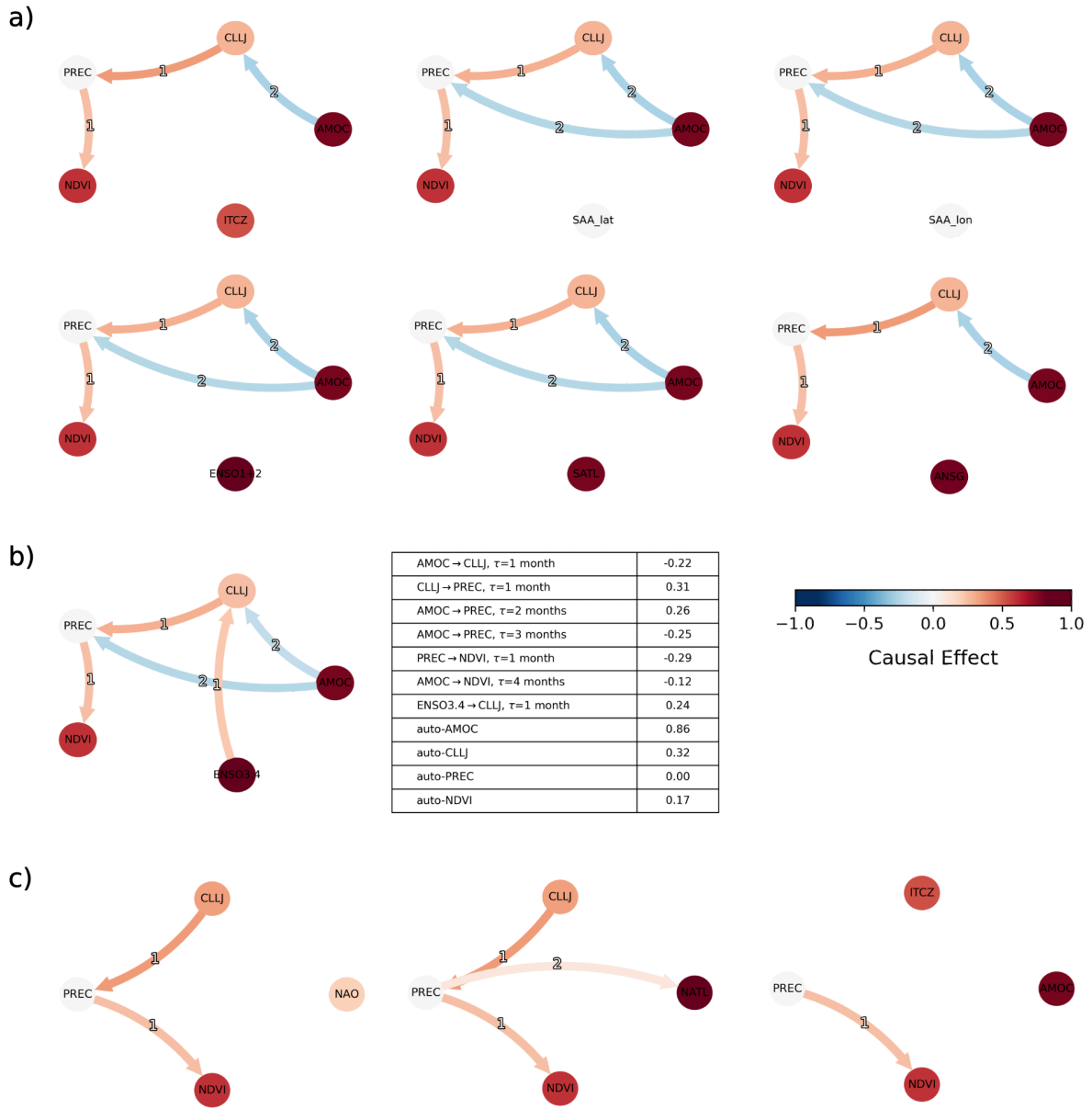

**Figure S2. AMOC→Southern AR dry season causal effect network with a) additional variables** (see Table 1), **b) ENSO3.4** shown with its CE table. **c) replacement of single variables** through related indices. The AMOC index is replaced with the NAO and the NATL, the CLLJ is replaced with the ITCZ index.

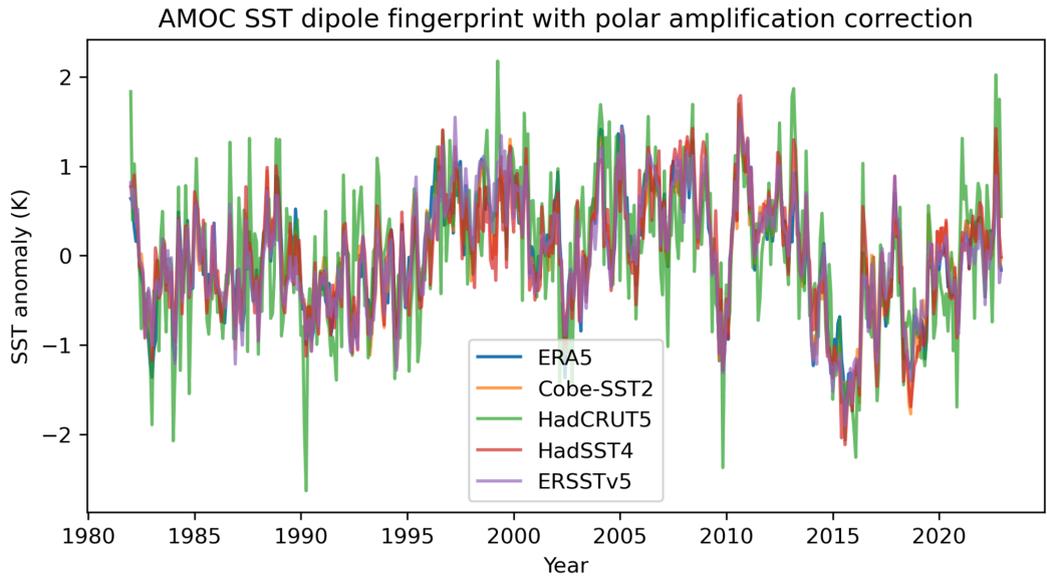

**Figure S3. AMOC SST dipole index from different data sources.** As Figure S1, with the AMOC$_{dipole}$ index adapted from Pontes & Menviel (Pontes and Menviel 2024).

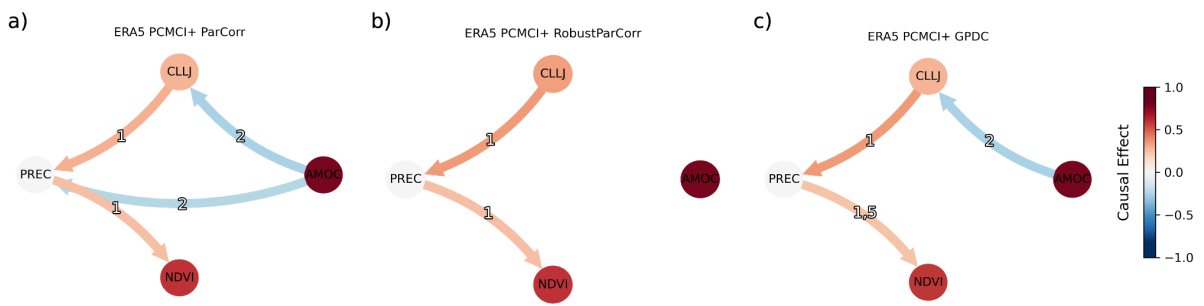

**Figure S4. AMOC→Southern AR dry season causal effect network with different conditional independence tests.** The causal network between AMOC index, CLLJ, and Southern AR precipitation and NDVI using different conditional independence tests in the causal discovery: a) with partial correlation (ParCorr), b) with partial correlation and the respective distributions of all variables Gaussianised prior to the independence testing (RobustParCorr), and c) with a non-linear Gaussian processes and distance (GPDC) based correlation.

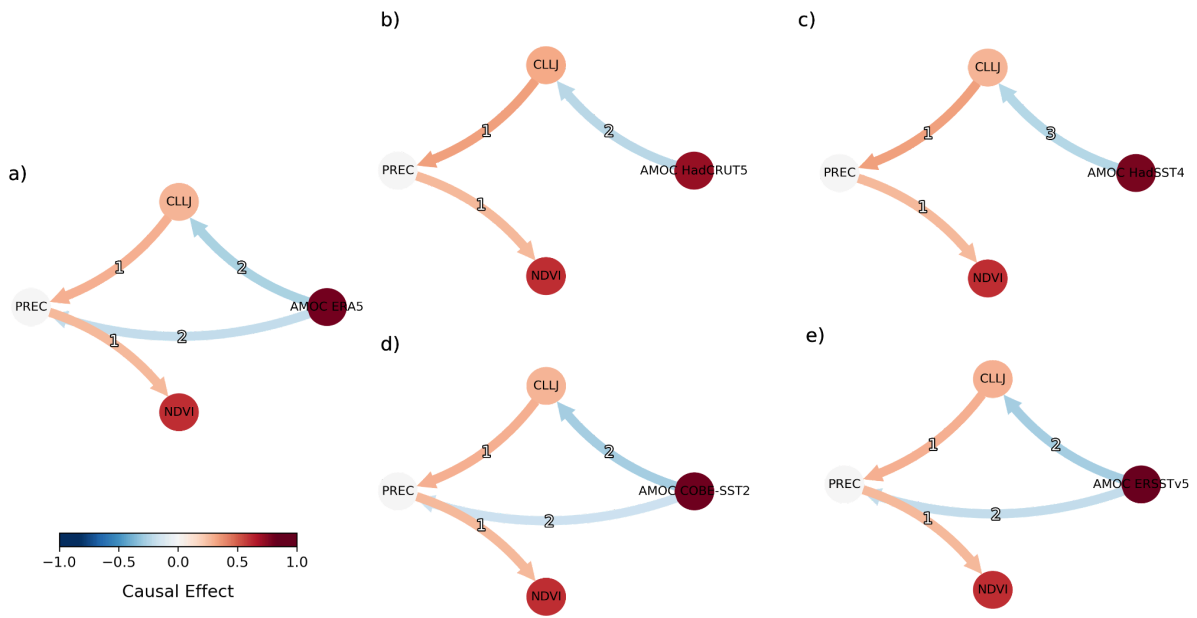

**Figure S5. AMOC→Southern AR dry season causal effect network using alternative data sources.** Here shown using GPCC data (Schneider *et al* 2022) as an alternative source for precipitation and a) ERA5, b) COBE-SST2, c) HadCRUT5, d) HadSST4, and e) NOAA ERSSTv5 as sources for SST.

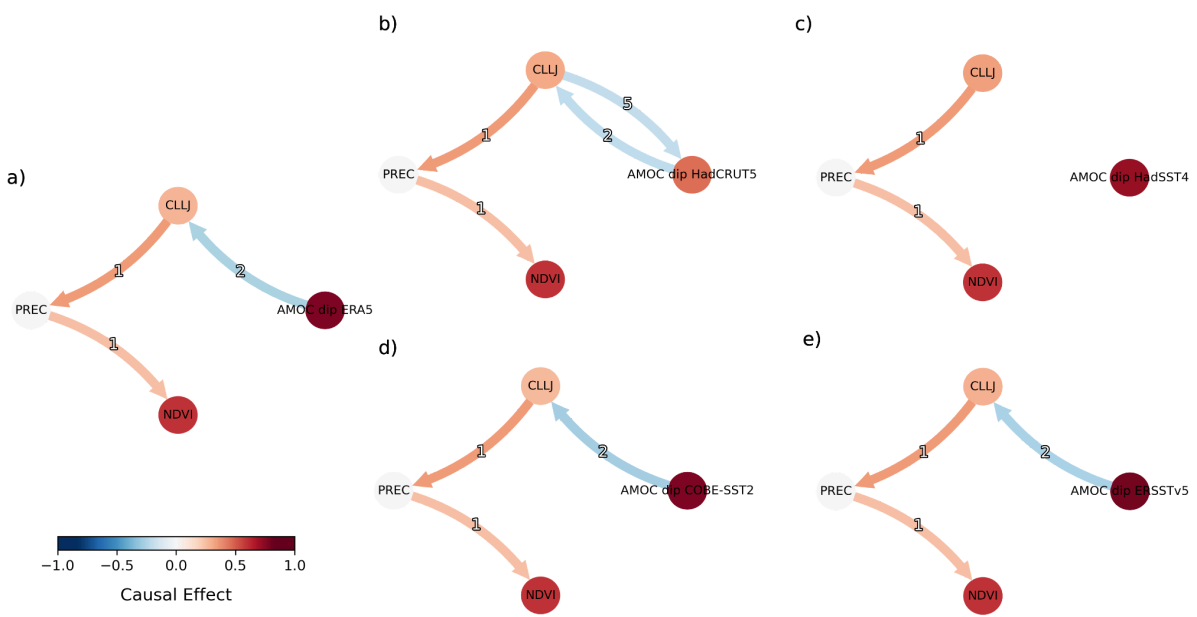

**Figure S6. AMOC→Southern AR dry season causal effect network with the AMOC$_{dipole}$ index.** As Figure S5, here using ERA5 precipitation and the AMOC$_{dipole}$ index based on different SST data sources.

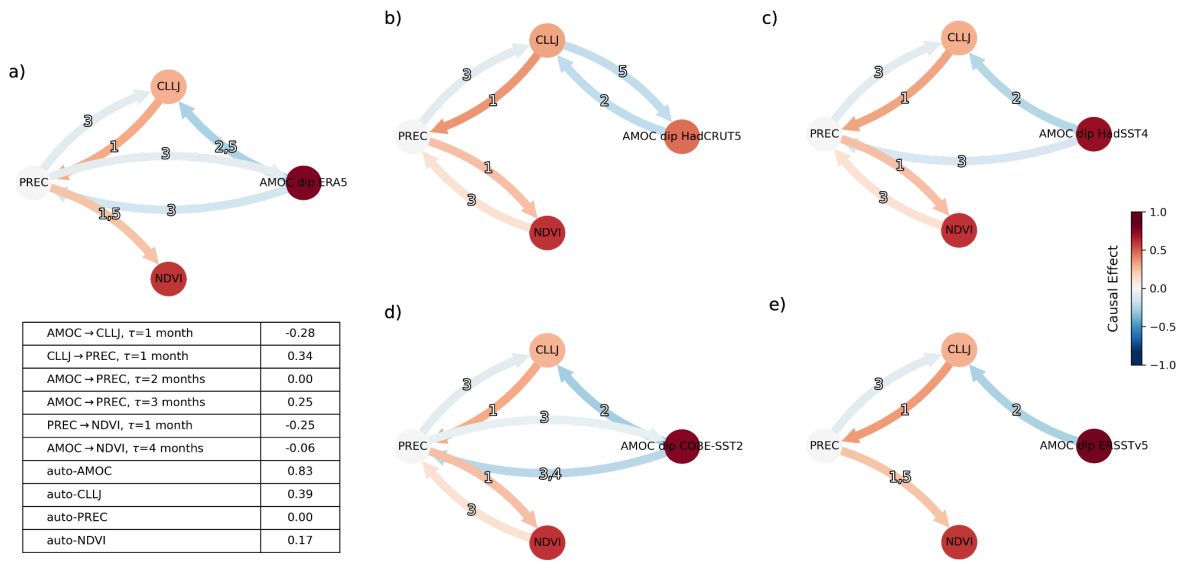

**Figure S7. AMOC→Southern AR dry season causal effect network with the AMOC$_{dipole}$ index.** As Figure S6, here with the non-linear GPDC independence test. We show the CE strength for a) evaluating the same links as in Figure 2.

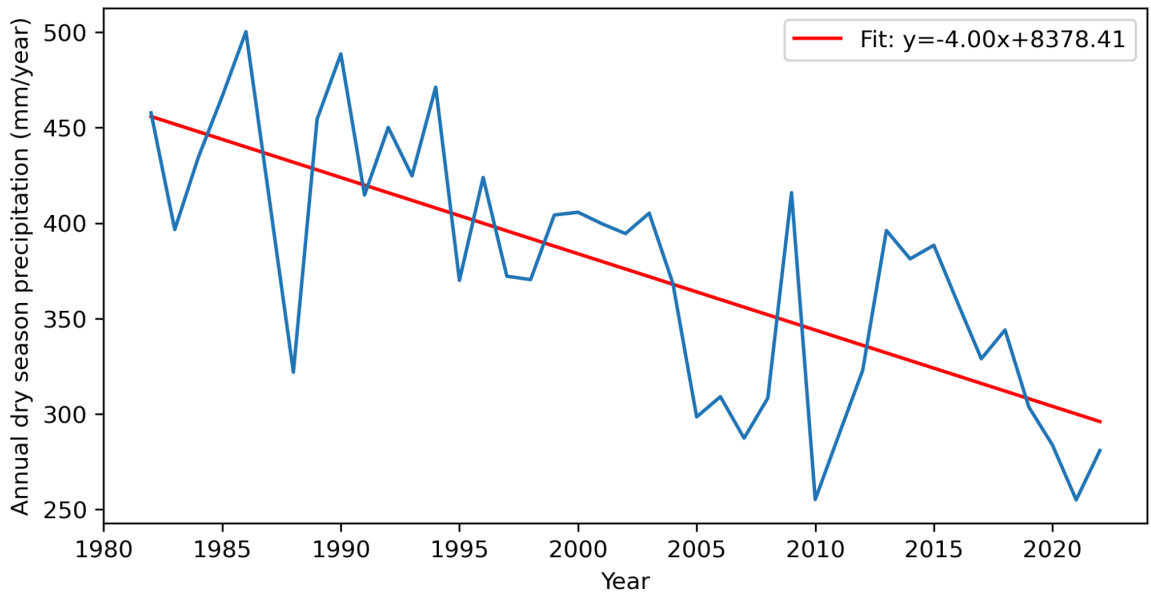

**Figure S8. Annual Southern AR dry season precipitation** (mm/year) from ERA5, with a linear fit (red). The dry season is defined as May-September.

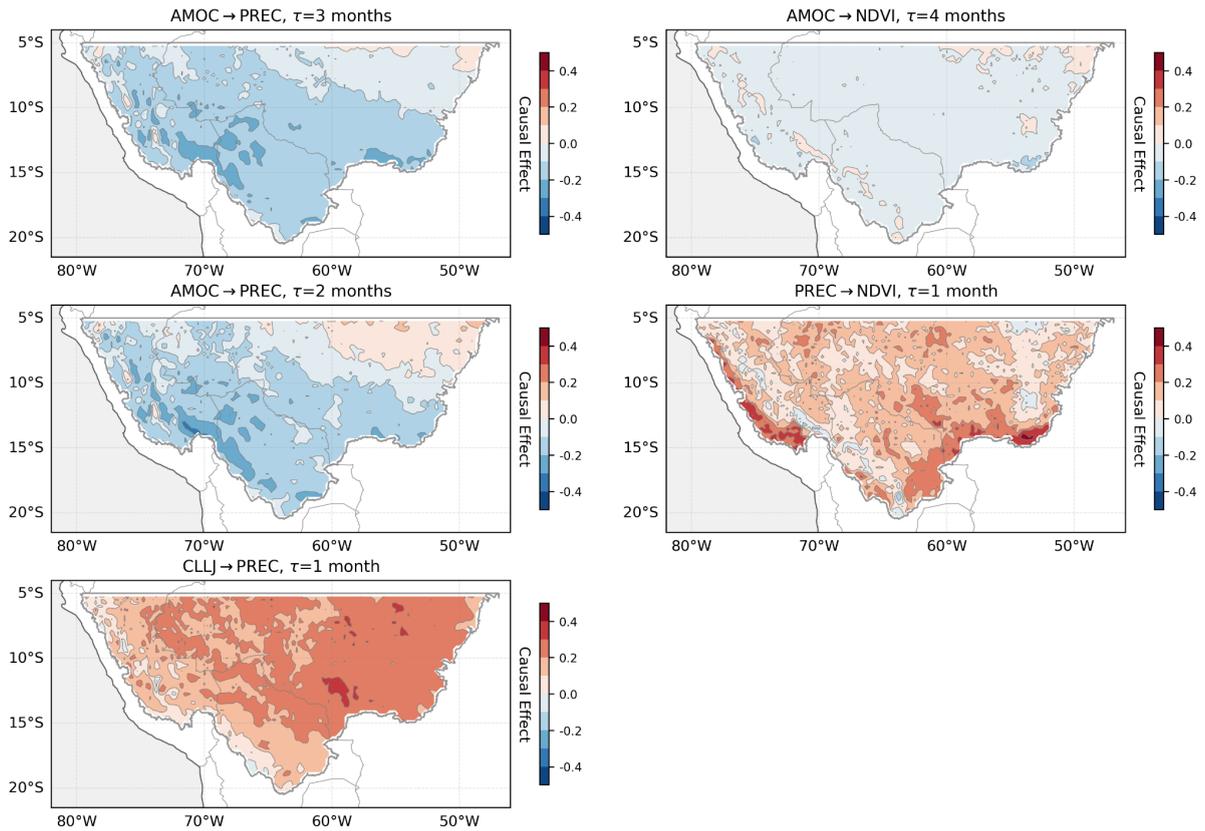

**Figure S9. Causal map for HadCRUT5 SST data.** Same as Figure. 3 in the main text but for HadCRUT5 SST data.

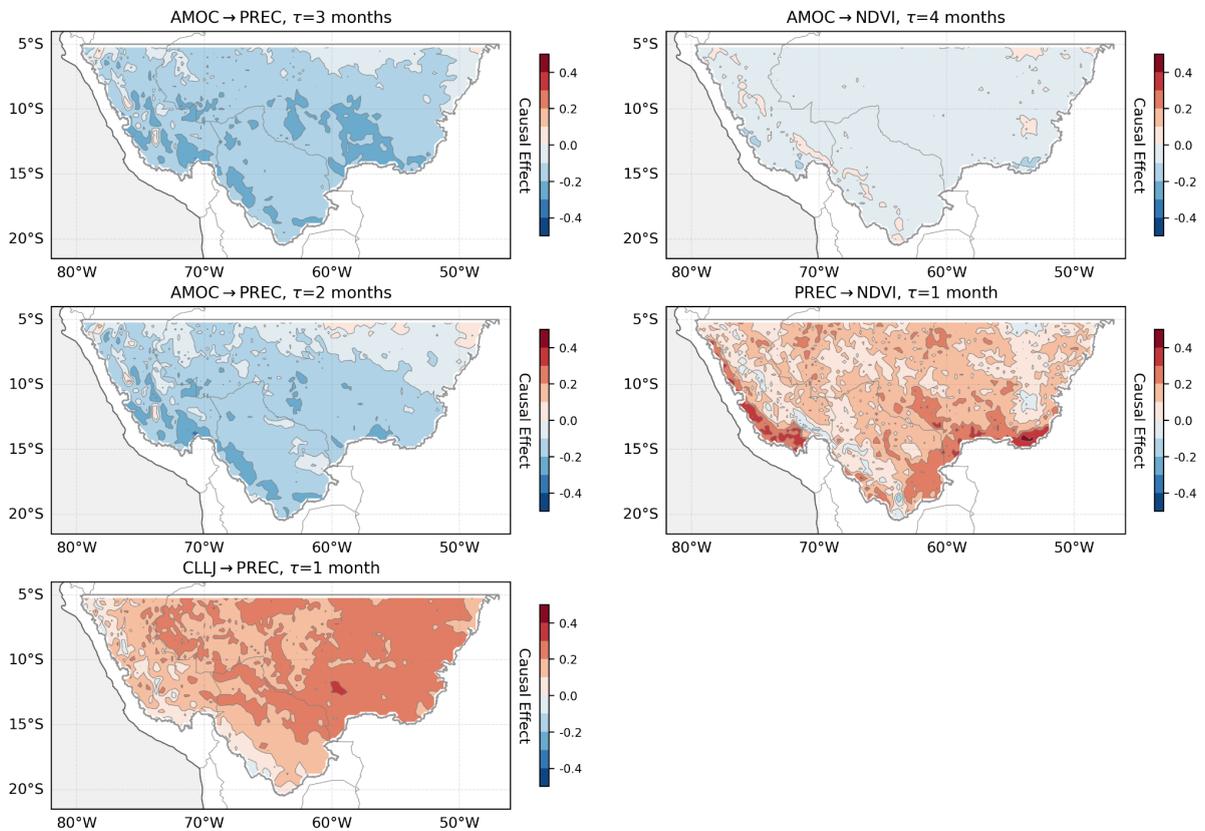

**Figure S10. Causal map for HadSST4 SST data.** Same as Figure 3 in the main text but for HadSST4 SST data.

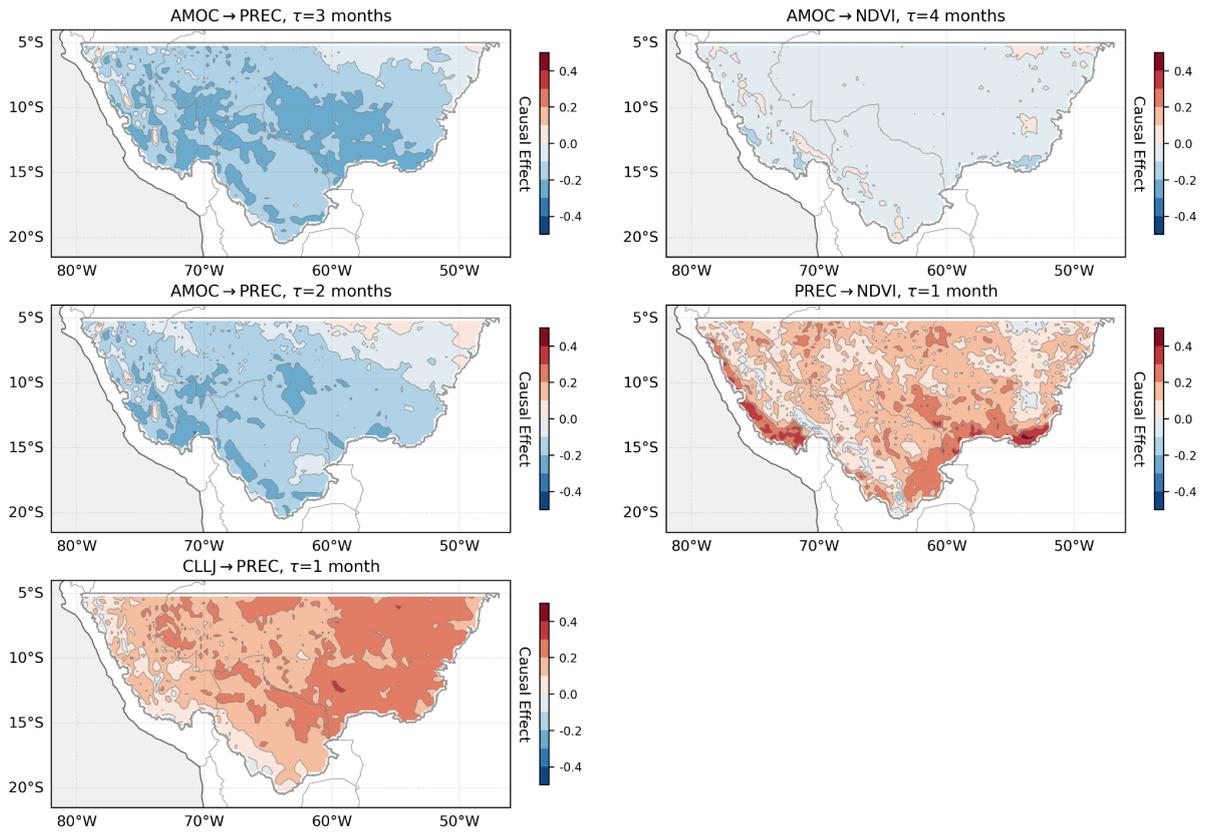

**Figure S11. Causal map for Cobe-SST2 SST data.** Same as Figure 3 in the main text but for COBE-SST2 SST data.

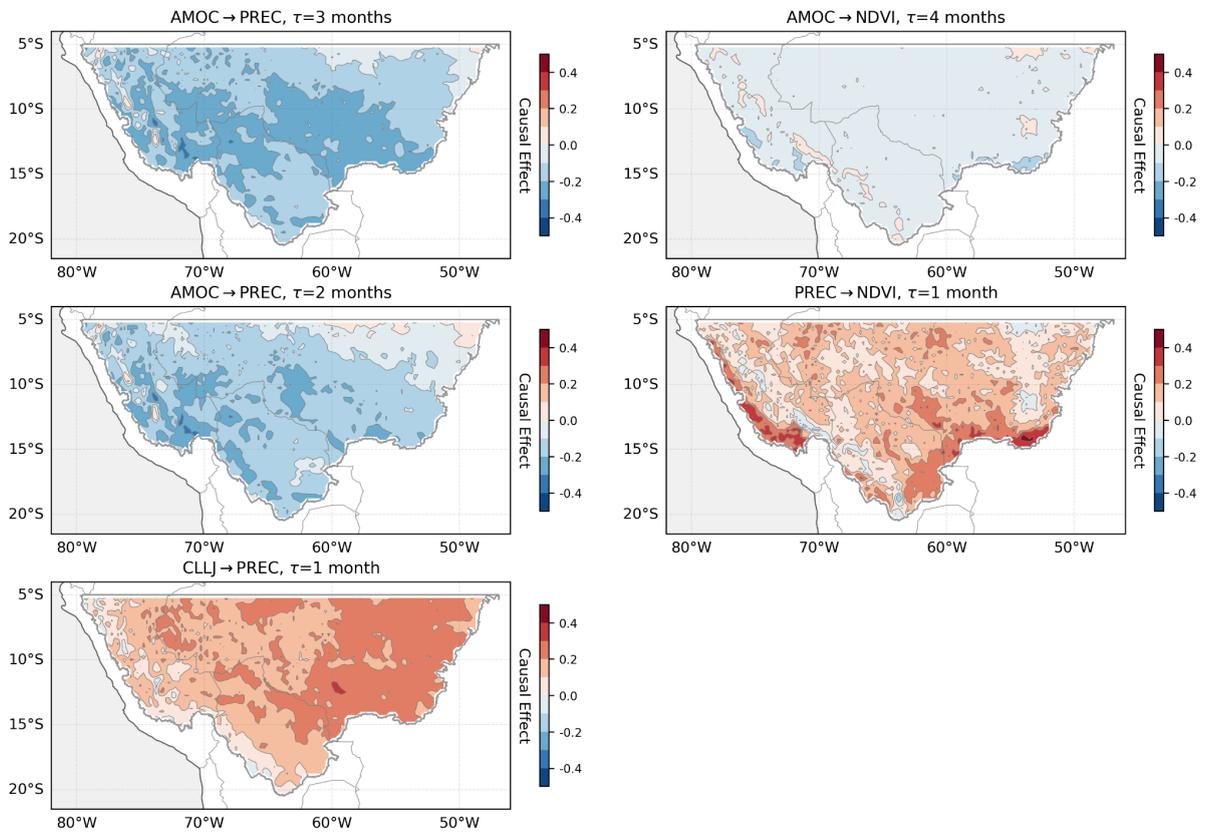

**Figure S12. Causal map for ERSSTv5 SST data.** Same as Figure 3 in the main text but for ERSSTv5 SST data.

27